\documentclass{article}

\title{Multielectronic Atom in Magnetic Field Revisited}
\author{O. Chavoya-Aceves\\
Camelback H. S., Phoenix, Arizona, USA.\\
chavoyao@yahoo.com}

\begin{document}
\maketitle

\begin{abstract}
The motion of a multi-electronic atom in an external
electro-magnetic field is reconsidered. We prove that according to
classical mechanics and electrodynamics, the assumption that the
interaction with the magnetic field is described by means of a
potential energy is no valid, and the trajectory of the center of
mass can be deflected by a magnetic field, even if the internal
angular momentum is zero. The characteristic equation of the
corresponding hamiltonian is not separable in three degrees of
freedom for the hydrogen atom.

\textbf{PACS 03.53.-w} Quantum Mechanics, atom in crossed fields
\end{abstract}
\section{Introduction}
In a previous paper \cite{OCHAVOYA-3} we showed that the
correspondence principle implies that the operator of angular
momentum for a particle in an electromagnetic field is:
\begin{equation}\label{operadores de momento angular}
  \hat{L}=\vec{r}\times(-i\hbar \nabla -\frac{q}{c}\vec{A}),
\end{equation}
where $\vec{A}$ is the vector potential. As we mentioned there,
(\ref{operadores de momento angular}) is required to guarantee
that the expected values of angular momentum are gauge-invariant.

Also, from the general relation
\begin{equation}\label{operador derivada}
  \frac{d{\hat{f}}}{dt}=\frac{i}{\hbar}[\hat{H},\hat{f}]+\frac{\partial \hat{f}}{\partial
  t},
\end{equation}
we realized that the term $-\frac{q}{c}\vec{r}\times\vec{A}$ has
to be included as part of the angular momentum, if the correct
contribution of the electric field to the torque is going to be
obtained---because
\[
\vec{E}=-\nabla V-\frac{1}{c}\frac{\partial \vec{A}}{\partial t},
\]
where $V$ is the scalar potential.

From this we concluded that the theory of angular momentum and the
theory of \emph{spin}, in particular, had to be revised.

For example, in view of (\ref{operadores de momento angular}), the
characteristic equation for the $z$ component of the angular
momentum in presence of a magnetic field $\vec{H}=H_0\hat{k}$,
that can be obtained from the vector potential
\[
\vec{A}=\frac{1}{2}\vec{H}\times\vec{r}
\]
is
\[
-i\hbar\frac{\partial \psi}{\partial \phi}-\frac{q}{c}H
r^2\sin^2(\theta)\psi=m\hbar\psi,
\]
that has not mono-valuated eigenfunctions, showing how the
interpretation of the magnetic and azimuthal quantum
numbers---that label the eigenfunctions of the operator
$-i\hbar\vec{r}\times\nabla$---is disrupted by the presence of the
magnetic field.

According to the axioms of quantum mechanics, the only values that
an observable can assume are the eigenvalues of its operator. If
this is true, the component of the angular momentum along the
magnetic field has not allowed values, neither the corresponding
component of the magnetic moment.

In another paper \cite{OCHAVOYA-4}, we reconsidered the hydrogen
atom in presence of an external magnetic field. The first thing we
noticed was that, following a classical lagrangian approach, it
can be proved that the motion of the center of mass and the
internal motion are not physically independent, and the classical
trajectory of the center of mass can be deflected if the field is
inhomogeneous, no matter if the internal angular momentum is zero;
an effect that is also predicted by Schr\"odinger theory, in view
of Ehrenfest's theorem---challenging the common belief about the
function of the Stern-Gerlach apparatus, as resolving the
eigen-states of an intrinsic angular momentum.

Also, we saw that the main evidence we have of the failure of
Schr\"odinger's theory to explain the properties of atoms in
presence of magnetic fields is not completely reliable, because
the usual formulation of the problem \cite[pp. 71 \& 541]{MESSIA}
\cite[pp. 359-60]{BOHM} is not accurate.

We concluded that there is a basis for an explanation of the
phenomena associated to \emph{spin} as consequences of the Laws of
Electrodynamics  as applied to systems of electrical charges as
wholes, but not as manifestations of intrinsic properties of
punctual particles, as was also sustained in a different way by
Bohr, who believed that the \emph{spin} was only an abstraction,
useful to compute the angular momentum\cite{GARRAWAY-STENHOLM}.

In this paper we obtain similar and other results for
multi-electronic atoms. We show that the assumption that the
interaction of a multi-electronic atom with an external magnetic
field can be fully described by means of a potential energy
\[
\Phi=\frac{e}{2 m_e c}\vec{L}\cdot\vec{H},
\]
where $\vec{L}$ is the internal angular momentum, is incompatible
with classical mechanics and electrodynamics.

We prove also that the separation of the motion of the center of
mass and the internal motion---neglecting quadratic terms in the
magnetic field---leads to what is known as the hamiltonian in
orthogonal crossed fields that is not separable \cite{MAIN} and
has been proved to have \emph{monodromy}: ``A dynamical property
that makes a global definition of angle-action variables and of
quantum numbers impossible''\cite{GUSHMAN}\cite{SADOVSKI}, at
least in the case of the hydrogen atom.

\section{On the Motion of a Neutral System in an External Electromagnetic Field}
We consider a system made of $N$ charged particles, with masses
$m_i$ and charges $q_i$ ($i=1,\cdots,N$), under the action of an
external time independent electromagnetic field, derivable from
the potentials $V_0(\vec{r})$ and $\vec{A}(\vec{r})$:
\begin{equation}\label{external field}
\vec{E}_{0}=-\nabla\vec{V_0},\ \ \ \vec{H}_{0}=\nabla\times\vec{A}
\end{equation}
We'll suppose that the system as a whole is electrically neutral,
in such way that:
\begin{equation}\label{condicion de neutralidad}
\sum_{i=1}^N q_i =0
\end{equation}
We will also neglect internal magnetic interactions, and write the
Lagrange's function of the system as:
\begin{equation}\label{lagrangian}
L=K-V(\vec{r}_1,\cdots,\vec{r}_N)+\sum_{i=1}^N q_i\left(
-V_0(\vec{r}_i)+\frac{\vec{v}_i}{c}\cdot\vec{A}(\vec{r}_i)\right).
\end{equation}
Here
\begin{equation}\label{energia interna}
  V(\vec{r}_1,\cdots,\vec{r}_N)=\sum_{i<j}\frac{q_iq_j}{\|\vec{r}_i-\vec{r}_j\|}.
\end{equation}

Our last assumption is that the external electromagnetic field
remains almost constant across the system and, therefore, that,
\emph{inside} the system, the corresponding potentials can be
reasonably approximated by linear functions.

Let $\vec{R}$ be the position vector of the center of mass and:
\begin{equation}\label{coordenadas internas}
\vec{r}_i=\vec{R}+\vec{\rho}_i
\end{equation}
The kinetic energy of the system is
\begin{equation}\label{energia cinetica}
  K=\frac{1}{2}M\dot{\vec{R}}^2+\frac{1}{2}\sum_{i=1}^N m_i
  \dot{\vec{\rho}}_i^2,
\end{equation}
and the internal electrical energy is easily written as a function
of the vectors $\vec{\rho}_i$:
\[
V(\vec{\rho}_1,\cdots,\vec{\rho}_N)=\sum_{i<j}\frac{q_iq_j}{\|\vec{\rho}_i-\vec{\rho}_j\|},
\]
whilst the other terms are replaced by their first-order
approximations.

Our last assumption implies that
\[
V_0(\vec{r}_i)\approx
V_0(\vec{R})+\nabla_{\vec{R}}V_0\cdot\vec{\rho}_i=%
V_0(\vec{R})-\vec{E}_0(\vec{R})\cdot\vec{\rho}_i.
\]

From this and (\ref{condicion de neutralidad}) we get
\begin{equation}\label{aproximacion de potencial externo}
  \sum_{i=1}^N q_i V_0(\vec{r}_i)\approx
  -\vec{P}\cdot\vec{E}_0(\vec{R}),
\end{equation}
where
\begin{equation}\label{electrical dipole}
  \vec{P}=\sum_{i=1}^N q_i \vec{\rho}_i,
\end{equation}
is the electrical dipole. As to the other terms, we have:
\begin{equation}\label{aproximacion de los terminos magneticos}
  \vec{v}_i\cdot\vec{A}(\vec{r}_i)=\vec{v}_i\cdot\vec{A}(\vec{R}+\vec{\rho}_i)%
  \approx\vec{A}(\vec{R})\cdot\vec{v}_i+((\vec{v}_i\cdot\nabla_{\vec{R}})\vec{A}(\vec{R})+%
  \vec{v}_i\times\vec{H}_0(\vec{R}))\cdot\vec{\rho}_i.
\end{equation}

Further:
\begin{equation}\label{primer termino magnetico}
  \sum_{i=1}^N\frac{q_i}{c}\vec{A}(\vec{R})\cdot\vec{v}_i=%
  \frac{1}{c}\left(\sum_{i=1}^Nq_i\right)\vec{A}(\vec{R})\cdot\dot{\vec{R}}%
  +\frac{1}{c}\vec{A}(\vec{R})\cdot\sum_{i=1}^Nq_i\dot{\vec{\rho}}_i=
  \frac{1}{c}\vec{A}(\vec{R})\cdot\dot{\vec{P}};
\end{equation}
\begin{equation}\label{segundo termino magnetico}
  \sum_{i=1}^N\frac{q_i}{c}((\vec{v}_i\cdot\nabla_{\vec{R}})\vec{A}(\vec{R}))\cdot\vec{\rho}_i=%
  \frac{1}{c}((\dot{\vec{R}}\cdot\nabla_{\vec{R}})\vec{A}(\vec{R}))\cdot\vec{P}+
  \frac{1}{c}\sum_{i=1}^Nq_i((\dot{\vec{\rho}}_i\cdot\nabla_{\vec{R}})\vec{A}(\vec{R}))\cdot\vec{\rho}_i
\end{equation}
\[
=\frac{1}{c}\dot{\vec{A}}(\vec{R})\cdot\vec{P}+%
\frac{1}{c}\sum_{i=1}^Nq_i((\dot{\vec{\rho}}_i\cdot\nabla_{\vec{R}})\vec{A}(\vec{R}))\cdot\vec{\rho}_i;
\]
\begin{equation}\label{tercer termino magnetico}
  \sum_{i=1}^N\frac{q_i}{c}(\vec{v}_i\times\vec{H}_0(\vec{R}))\cdot\vec{\rho}_i=%
  \sum_{i=1}^N\frac{q_i}{c}(\vec{\rho}_i\times\vec{v}_i)\cdot\vec{H}_0(\vec{R})
\end{equation}
\[
=\vec{H}_0(\vec{R})\cdot\sum_{i=1}^N\frac{q_i}{c}\vec{\rho}_i\times\dot{\vec{R}}+%
\vec{H}_0(\vec{R})\cdot\sum_{i=1}^N\frac{q_i}{c}\vec{\rho}_i\times\dot{\vec{\rho}}_i.
\]

Now we can write our first approximation to the Lagrange's
Function:
\begin{equation}\label{primera aproximacion de la lagrangiana}
  L(\vec{R},\vec{\rho}_1,\cdots,\vec{\rho}_N,\dot{\vec{\rho}}_1,\cdots,\dot{\vec{\rho}}_N)=%
  \frac{1}{2}M\dot{\vec{R}}^2+\frac{1}{2}\sum_{i=1}^Nm_i\dot{\vec{\rho}}_i^2-V+\vec{P}\cdot\vec{E}_0(\vec{R})+
\end{equation}
\[
\frac{1}{c}\sum_{i=1}^Nq_i((\dot{\vec{\rho}}_i\cdot\nabla_{\vec{R}})\vec{A}(\vec{R}))\cdot\vec{\rho}_i+\vec{H}_0(\vec{R})\cdot\sum_{i=1}^N\frac{q_i}{c}\vec{\rho}_i\times\dot{\vec{R}}+%
\vec{H}_0(\vec{R})\cdot\sum_{i=1}^N\frac{q_i}{c}\vec{\rho}_i\times\dot{\vec{\rho}}_i%
\]
where the terms
\[
\frac{1}{c}\vec{A}(\vec{R})\cdot\dot{\vec{P}}+\frac{1}{c}\dot{\vec{A}}(\vec{R})\cdot\vec{P}
\]
have been omitted, because they add up to a total derivative.

Let's consider a typical summand in $
\sum_{i=1}^Nq_i((\dot{\vec{\rho}}_i\cdot\nabla_{\vec{R}})\vec{A}(\vec{R}))\cdot\vec{\rho}_i
$:
\[
((\dot{\vec{\rho}}\cdot\nabla_{\vec{R}})\vec{A}(\vec{R}))\cdot\vec{\rho}=%
\rho_i\dot{\rho}_j\partial_jA_i=\frac{\rho_i\dot{\rho}_j-\rho_j\dot{\rho}_i}{2}\partial_jA_i+\frac{\rho_i\dot{\rho}_j+\rho_j\dot{\rho}_i}{2}\partial_jA_i=
\]
\[
\frac{\epsilon_{ijk}\epsilon_{kab}}{2}\rho_a\dot{\rho}_b\partial_jA_i+\frac{\rho_i\dot{\rho}_j+\rho_j\dot{\rho}_i}{2}\partial_jA_i=-\frac{1}{2}\vec{H}(\vec{R})\cdot(\vec{\rho}\times\dot{\vec{\rho}})+
\]
\[
\frac{1}{2}(\vec{\rho}\cdot((\dot{\vec{\rho}}\cdot\nabla_{\vec{R}})\vec{A})+\dot{\vec{\rho}}\cdot((\vec{\rho}\cdot\nabla_{\vec{R}})\vec{A}))
\]

The last term can be shown to be equal to a total-derivative plus
a term that includes second derivatives of the external vector
potential, that can be neglected since we have supposed that,
\emph{inside} the system, the external potentials can be
reasonably approximated by linear functions. On this basis, we
substitute the Lagrange's Function by
\begin{equation}\label{segunda aproximacion de la lagrangiana}
  L=\frac{1}{2}M\dot{\vec{R}}^2+%
  \frac{1}{2}\sum_{i=1}^Nm_i\dot{\vec{\rho}}_i^2-V+%
  \vec{P}\cdot\left(\vec{E}_0(\vec{R})+%
  \frac{\dot{\vec{R}}}{c}\times\vec{H}_0(\vec{R})\right)+%
  \frac{1}{2}\vec{H}_0(\vec{R})\cdot\sum_{i=1}^N\frac{q_i}{c}\vec{\rho}_i\times\dot{\vec{\rho}}_i%
\end{equation}

Given that the vectors $\vec{\rho}_i$ are subject to the condition
\begin{equation}\label{condicion de restriccion}
  \sum_{i=1}^N m_i\vec{\rho}_i,
\end{equation}
to obtain the equations of motion, we have to introduce a time
dependent vectorial multiplier, which leads to the modified
Lagrange's Function:
\begin{equation}\label{tercera aproximacion de la lagrangiana}
  L'=\frac{1}{2}M\dot{\vec{R}}^2+%
  \frac{1}{2}\sum_{i=1}^Nm_i\dot{\vec{\rho}}_i^2-V+%
  \vec{P}\cdot\left(\vec{E}_0(\vec{R})+%
  \frac{\dot{\vec{R}}}{c}\times\vec{H}_0(\vec{R})\right)+%
\end{equation}
\[
\frac{1}{2}\vec{H}_0(\vec{R})\cdot\sum_{i=1}^N\frac{q_i}{c}\vec{\rho}_i\times\dot{\vec{\rho}}_i+\vec{\lambda}\cdot\sum_{i=1}^nm_i\vec{\rho}_i
\]

The equations of motion are:
\begin{equation}\label{ecuacion para el centro de masas}
  \frac{d}{dt}\left(M\dot{\vec{R}}+\frac{1}{c}\vec{H}_0\times\vec{P}\right)=%
  (\vec{P}\cdot\nabla_{\vec{R}})\vec{E}_0(\vec{R})+
\end{equation}
\[
\left(\left(\frac{1}{c}\vec{P}\times\dot{\vec{R}}+\frac{1}{2c}\sum_{i=1}^N
q_i\vec{\rho}_i\times\dot{\vec{\rho}}_i\right)\cdot\nabla_{\vec{R}}\right)\vec{H}_0(\vec{R})
\]
and
\begin{equation}\label{ecuaciones del movimiento interno}
  \frac{d}{dt}\left(m_i\dot{\vec{\rho}}_i+\frac{q_i}{2c}\vec{H}_0(\vec{R})\times\vec{\rho}_i\right)=%
  -\frac{\partial V}{\partial \vec{\rho}_i}+
  q_i\left(
  \vec{E}_0(\vec{R})+\frac{\dot{\vec{R}}}{c}\times\vec{H}_0(\vec{R})\right)
\end{equation}
\[
+\frac{q_i}{2c}\dot{\vec{\rho}}_i\times\vec{H}_0(\vec{R})+m_i\vec{\lambda}
\]

Considering that
\[
\frac{d}{dt}(\vec{H}_0\times\vec{P})=%
((\dot{\vec{R}}\cdot\nabla_{\vec{R}})\vec{H}_0)\times\vec{P}%
+\vec{H}_0\times\dot{\vec{P}}=
\]
\[
\nabla_{\vec{R}}(\dot{\vec{R}}\cdot\vec{H})\times\vec{P}+\vec{H}_0\times\dot{\vec{P}}
\]
and
\[
\frac{d}{dt}(\vec{H}_0\times\vec{\rho}_i)=
\nabla_{\vec{R}}(\dot{\vec{R}}\cdot\vec{H})\times\vec{\rho}_i+\vec{H}_0\times\dot{\vec{\rho}}_i
\]
(Where we have used the identity
\[
\nabla (\vec{a}\cdot\vec{b})=%
(\vec{a}\cdot\nabla)\vec{b}+%
(\vec{b}\cdot\nabla)\vec{a}+%
\vec{a}\times(\nabla\times\vec{b})+%
\vec{b}\times(\nabla\times\vec{a})
\]
and the fact that
\[
\nabla_{\vec{R}}\times\vec{H}_0(\vec{R})=\vec{0},
\]
since $\vec{H}_0$ is an \emph{external} field.), eqs.
(\ref{ecuacion para el centro de masas} \& \ref{ecuaciones del
movimiento interno}) can be rewritten as:
\begin{equation}\label{segunda ecuacion para el centro de masas}
  M\ddot{\vec{R}}=%
  (\vec{P}\cdot\nabla_{\vec{R}})\vec{E}_0(\vec{R})+%
  \frac{1}{c}\dot{\vec{P}}\times\vec{H}_0(\vec{R})+%
  \frac{1}{c}\vec{P}\times\nabla_{\vec{R}}(\dot{\vec{R}}\cdot\vec{H}_0(\vec{R}))+%
\end{equation}
\[
\left(\left(\frac{1}{c}\vec{P}\times\dot{\vec{R}}+\frac{1}{2c}\sum_{i=1}^N
q_i\vec{\rho}_i\times\dot{\vec{\rho}}_i\right)\cdot\nabla_{\vec{R}}\right)\vec{H}_0(\vec{R})
\]
and
\begin{equation}\label{segunda ecuacion para el movimiento interno}
  m_i\ddot{\vec{\rho}}_i=-\frac{\partial V}{\partial\vec{\rho}_i}+%
  q_i\left(\vec{E}_0(\vec{R})+\frac{\dot{\vec{R}}}{c}\times\vec{H}_0(\vec{R})\right)+%
\end{equation}
\[
\frac{q_i}{2c}\vec{\rho}_i\times\nabla_{\vec{R}}(\dot{\vec{R}}\cdot\vec{H}_0(\vec{R}))+%
  \frac{q_i}{c}\dot{\vec{\rho}}_i\times\vec{H}_0(\vec{R})+m_i\vec{\lambda}
\]

From (\ref{condicion de neutralidad}), (\ref{segunda ecuacion para
el movimiento interno})  and the relations
\[
\sum_{i=1}^n m_i\vec{\rho}_i=\vec{0}, \ \ \  \sum_{i=1}^n
\frac{\partial V}{\partial \vec{\rho}_i}=\vec{0},
\]
we get:
\begin{equation}\label{calculo del multiplicador}
  \vec{\lambda}=-\frac{\vec{P}}{2Mc}\times\nabla_{\vec{R}}(\dot{\vec{R}}\cdot\vec{H}_0(\vec{R}))-%
  \frac{\dot{\vec{P}}}{Mc}\times\vec{H}_0(\vec{R}).
\end{equation}
Therefore
\begin{equation}\label{tercera ecuacion para el movimiento interno}
  m_i\ddot{\vec{\rho}}_i=-\frac{\partial V}{\partial\vec{\rho}_i}+%
  q_i\left(\vec{E}_0(\vec{R})+\frac{\dot{\vec{R}}}{c}\times\vec{H}_0(\vec{R})\right)+%
\end{equation}
\[
\frac{1}{2c}(q_i\vec{\rho}_i-\frac{m_i}{M}\vec{P})\times%
\nabla_{\vec{R}}(\dot{\vec{R}}\cdot\vec{H}_0(\vec{R}))+%
\frac{1}{c}(q_i\dot{\vec{\rho}}_i-\frac{m_i}{M}\dot{\vec{P}})%
\times\vec{H}_0(\vec{R})
\]
\section{Multi-Electronic Atom in a Magnetic Field}
Equations (\ref{segunda ecuacion para el centro de masas} \&
\ref{tercera ecuacion para el movimiento interno}) clearly show
that the internal motion and the motion of the center of mass are
not independent in presence of an external electromagnetic field,
as was supposed to formulate the quantum description of
multi-electronic atoms.

In case of a purely magnetic external field, the equation of
motion of the center of mass is simplified to:
\begin{equation}\label{ecuacion para el centro de masas-magnetico}
  M\ddot{\vec{R}}=%
  \frac{1}{c}\dot{\vec{P}}\times\vec{H}_0(\vec{R})+%
  \frac{1}{c}\vec{P}\times\nabla_{\vec{R}}(\dot{\vec{R}}\cdot\vec{H}_0(\vec{R}))+%
\end{equation}
\[
\left(\left(\frac{1}{c}\vec{P}\times\dot{\vec{R}}+\frac{1}{2c}\sum_{i=1}^N
q_i\vec{\rho}_i\times\dot{\vec{\rho}}_i\right)\cdot\nabla_{\vec{R}}\right)\vec{H}_0(\vec{R})
\]
The term
\begin{equation}\label{termino de larmor}
  \left(\left(\frac{1}{2c}\sum_{i=1}^N
q_i\vec{\rho}_i\times\dot{\vec{\rho}}_i\right)\cdot\nabla_{\vec{R}}\right)\vec{H}_0(\vec{R})
\end{equation}
is what is usually substituted by
\begin{equation}\label{termino de larmor aproximado}
  -\frac{e}{2m_ec}(\vec{L}\cdot\nabla_{\vec{R}})\vec{H}_0(\vec{R}),
\end{equation}
which is valid only in case that all the particles have the same
charge-mass relation, which is not true for common atoms.

It's evident that, given that the mass of the nucleus is much
bigger than the electronic mass, we can neglect the corresponding
contribution in (\ref{termino de larmor}) to justify its
substitution by (\ref{termino de larmor aproximado}). But this is
just an approximation, and not a fundamental relation, as its
usually presented, to attribute to classical physics the naive
view that an atom in a magnetic field can be considered as a tiny
circuit. Even more, if this approximation is introduced,
(\ref{ecuacion para el centro de masas-magnetico}) is transformed
into:
\begin{equation}\label{ecuacion para el centro de masas-magnetico aproximado}
  M\ddot{\vec{R}}=%
  \frac{1}{c}\dot{\vec{P}}\times\vec{H}_0(\vec{R})+%
  \frac{1}{c}\vec{P}\times\nabla_{\vec{R}}(\dot{\vec{R}}\cdot\vec{H}_0(\vec{R}))%
  -\left(\left(\frac{e}{m_ec}\vec{S}+\frac{e}{2m_ec}\vec{L}\right)\cdot\nabla_{\vec{R}}\right)\vec{H}_0(\vec{R}),
\end{equation}
where
\[
\frac{e}{m_ec}\vec{S}=\frac{1}{c}\vec{P}\times\dot{\vec{R}}
\]

Equation (\ref{ecuacion para el centro de masas-magnetico
aproximado}) looks like the equation of motion of a particle with
an intrinsic angular momentum and by no means implies that ``The
trajectory of a multi-electronic atom cannot be deflected by an
inhomogeneous magnetic field if the internal angular momentum is
equal to zero.'' as it's claimed to demonstrate that the result of
the Stern-Gerlach experiment is not predicted by classical
mechanics and cannot be explained but by introducing \emph{spin}
variables. In fact, equation (\ref{ecuacion para el centro de
masas-magnetico}) includes the force
\[
\vec{ f}_{||}=\frac{1}{c}\vec{P}\times\nabla_{\vec{R}}(\dot{\vec{R}}\cdot\vec{H}_0(\vec{R}))%
\]
that appears when the component of the velocity along the magnetic
field is not negligible, which is not predicted by the usual
theory.

To obtain an approximated quantum representation of the motion,
we'll suppose that the trajectory of the nucleus coincides with
the trajectory of the center of mass, in such way that the
Lagrange's function can be approximated by:
\begin{equation}\label{lagrangiano aproximado magnetico}
  L=\frac{1}{2}M\dot{\vec{R}}^2+%
\frac{m_e}{2}\sum_{i=1}^Z\dot{\vec{\rho}}_i^2-V+%
\frac{\vec{P}}{c}\cdot(\dot{\vec{R}}\times\vec{H}(\vec{R}))-%
\frac{e}{2m_ec}\vec{L}\cdot\vec{H}_0(\vec{R})
\end{equation}
where $Z$ is the atomic number,
\[
V=-\sum_{i=1}^Z\frac{e^2}{\|\vec{R}-\vec{\rho}_i\|}+%
\sum_{i<j}\frac{e^2}{\|\vec{\rho}_i-\vec{\rho}_j\|},
\]
\[
\vec{P}=-e\sum_{i=1}^Z\vec{\rho}_i,
\]
and
\[
\vec{L}=m_e\sum_{i=1}^Z\vec{\rho}_i\times\dot{\vec{\rho}}_i
\]

The corresponding momenta are:
\begin{equation}\label{impulso generalizado del centro de masas}
  \vec{p}_{\vec{R}}=M\dot{\vec{R}}+\frac{1}{c}\vec{H}(\vec{R})\times\vec{P}
\end{equation}
\begin{equation}\label{impulso electronico generalizado}
  \vec{p}_{\vec{\rho}_i}=m_e\dot{\vec{\rho}}_i-\frac{e}{2c}\vec{H}_0(\vec{R})\times\vec{\rho}_i,
\end{equation}
from which we can get the energy, which is a constant of motion:
\begin{equation}\label{energy}
  E=\frac{1}{2}M\dot{\vec{R}}^2+\frac{m_e}{2}\sum_{i=1}^Z\dot{\vec{\rho}}_i^2+V,
\end{equation}

Equation (\ref{energy}) clearly shows  the unsoundness of the
assumption that in conformity with the theoretical framework of
classical mechanics and electrodynamics the interaction of an atom
with a magnetic field is completely described by means of the
potential energy
\begin{equation}\label{energia magnetica}
  \Phi=\frac{e}{2m_ec}\vec{L}\cdot\vec{H}.
\end{equation}

According to classical mechanics the change of potential energy is
equal to the work done by the corresponding forces and, according
to electrodynamics, the magnetic force is perpendicular to the
velocity and, therefore, does not work. That's why the energy
(\ref{energy}) does not include magnetic terms, just the kinetic
energy plus the electrical potential energy, as predicted by
electrodynamics, but the Hamilton's Function does:
\begin{equation}\label{hamilton function}
  H=\frac{(\vec{p}_{\vec{R}}-\frac{1}{c}\vec{H}_0(\vec{R})\times\vec{P})^2}{2M}+%
  \sum_{i=1}^Z\frac{(\vec{p}_{\vec{\rho}_i}+\frac{e}{2c}\vec{H}_0(\vec{R})\times\vec{\rho}_i)^2}{2m_e}+V.
\end{equation}

The Hamiltonian is:
\begin{equation}\label{hamiltoniano}
  \hat{H}=\frac{(-i\hbar\nabla_{\vec{R}}-\frac{1}{c}\vec{H}_0(\vec{R})\times\vec{P})^2}{2M}+%
  \sum_{i=1}^Z\frac{(-i\hbar\nabla_{\vec{\rho}_i}+\frac{e}{2c}\vec{H}_0(\vec{R})\times\vec{\rho}_i)^2}{2m_e}+V.
\end{equation}

After some algebra, considering that
\[
\nabla_{\vec{R}}\times\vec{H}_{\vec{R}}=\vec{0},
\]
the Hamiltonian is transformed into:
\begin{equation}\label{transformed hamiltonian}
 \hat{H}= -\frac{\hbar^2}{2M}\nabla_{\vec{R}}^2-%
  \frac{\hbar^2}{2m_e}\sum_{i=1}^Z\nabla_{{\rho}_i}^2+ V+%
  \frac{i\hbar}{Mc}(\vec{H}_0(\vec{R})\times\vec{P})\cdot\nabla_{\vec{R}}+%
\end{equation}
\[
  \frac{e}{2m_ec}\vec{H}_0(\vec{R})\cdot\hat{L}+
 \frac{1}{2Mc^2}(\vec{H}_0(\vec{R})\times\vec{P})^2+%
        \frac{e^2}{8m_ec^2}\sum_{i=1}^Z(\vec{H}_0(R)\times\vec{\rho}_i)^2,
\]
where
\[
\hat{l}=\sum_{i=1}^Z-i\hbar\vec{\rho}_i\times\nabla_{\vec{\rho}_i}
\]

If the last two terms can be neglected, the Hamiltonian is
simplified to
\begin{equation}\label{hamiltoniano aproximado}
 \hat{H}= -\frac{\hbar^2}{2M}\nabla_{\vec{R}}^2-%
  \frac{\hbar^2}{2m_e}\sum_{i=1}^Z\nabla_{{\rho}_i}^2+ V+%
  \frac{i\hbar}{Mc}(\vec{H}_0(\vec{R})\times\vec{P})\cdot\nabla_{\vec{R}}+%
\end{equation}
\[
  \frac{e}{2m_ec}\vec{H}_0(\vec{R})\cdot\hat{l}.
\]

Compare this operator to the operator in \cite[pp. 71 \&
541]{MESSIA} to see that the conclusion that the properties of
atoms in magnetic fields cannot be explained but through the
introduction of \emph{spin} variables is, at least, precipitated,
and probably wrong. To avoid come to the conclusion that, after
all, (\ref{energia magnetica}) represents a sort of magnetic
potential energy, notice that, in view of (\ref{impulso
generalizado del centro de masas}), the operators $-i\hbar\nabla$
in (\ref{hamiltoniano aproximado}) do not correspond to linear
momentum and the operator $\frac{e}{2m_ec}\hat{l}$ does not
correspond to the magnetic moment. If the missing term is
introduced in the Hamiltonian, as well as the spinorial
terms---that have the form (\ref{energia magnetica}) that we have
already discarded---the agreement of the predicted spectra with
the experimental results will be disrupted.

Let's suppose that the magnetic field is uniform and make the
substitution
\begin{equation}\label{separacion de la funcion de onda}
  \Psi(\vec{R},\vec{\rho}_1,\cdots,\vec{\rho}_Z)=e^{\frac{i}{\hbar}\vec{p}_{\vec{R}}\cdot\vec{R}}\psi(\vec{\rho}_1,\cdots,\vec{\rho}_Z)
\end{equation}

The characteristic equation for (\ref{hamiltoniano aproximado}) is
transformed into:
\begin{equation}\label{ecuacion de eigenvalores movimiento interno}
\left( -\frac{\hbar^2}{2m_e}\sum_{i=1}^Z\nabla_{{\rho}_i}^2+ V-%
  \vec{P}\cdot\frac{\vec{p}_{\vec{R}}\times\vec{H}_0}{Mc}+
    \frac{e}{2m_ec}\vec{H}_0\cdot\hat{L}\right)\psi=\epsilon\psi,%
\end{equation}
where
\[
\epsilon=E-\frac{{\vec{p}_{\vec{R}}}^{\ 2}}{2M}
\]

This is the characteristic equation of what is known as the
Hamiltonian in orthogonal crossed fields---$\vec{E}_0=\frac{\vec{p}_{\vec{R}}%
\times\vec{H}_0}{Mc}$ and $\vec{H}_0$---that is not separable in
three degrees of freedom, for the hydrogen atom, because the
presence of the electric field breaks the azimuthal
symmetry\cite{MAIN}. Furthermore, recently, it has been proved
that the hydrogen atom in orthogonal crossed fields has
\emph{monodromy}: ``A dynamical property that makes a global
definition of angle-action variables and of quantum numbers
impossible''\cite{{GUSHMAN}}\cite{SADOVSKI} which might explain
the need to introduce changes in Schr\"odinger's formalism.

\end{document}